\documentclass[aps,pre,twocolumn,superscriptaddress]{revtex4-1}
\usepackage{amsfonts}
\usepackage{graphicx}
\usepackage{epstopdf}
\usepackage{epsfig}
\usepackage{amsmath}
\usepackage{color}
\usepackage{makecell}
\usepackage{mathtools}

\begin{document}

\title{Hysteresis in anesthesia and recovery: Experimental observation and dynamical mechanism}

\author{Chun-Wang Su}
\affiliation{The Key Laboratory of Biomedical Information Engineering of Ministry of Education, Institute of Health and Rehabilitation Science, School of Life Science and Technology, Xi'an Jiaotong University, Xi'an, Shaanxi, 710049, P. R. China}
\affiliation{National Engineering Research Center for Healthcare Devices. Guangzhou, Guangdong, 510500, P.R. China}
\affiliation{School of Physical Science and Technology, Lanzhou University, Lanzhou, 730000, P.R. China}

\author{Liang Zheng}
\affiliation{The Key Laboratory of Biomedical Information Engineering of Ministry of Education, Institute of Health and Rehabilitation Science, School of Life Science and Technology, Xi'an Jiaotong University, Xi'an, Shaanxi, 710049, P. R. China}
\affiliation{National Engineering Research Center for Healthcare Devices. Guangzhou, Guangdong, 510500, P.R. China}
\affiliation{The Key Laboratory of Neuro-informatics \& Rehabilitation Engineering of Ministry of Civil Affairs, Xi'an, Shaanxi, 710049, P. R. China}

\author{You-Jun Li}
\affiliation{The Key Laboratory of Biomedical Information Engineering of Ministry of Education, Institute of Health and Rehabilitation Science, School of Life Science and Technology, Xi'an Jiaotong University, Xi'an, Shaanxi, 710049, P. R. China}
\affiliation{National Engineering Research Center for Healthcare Devices. Guangzhou, Guangdong, 510500, P.R. China}
\affiliation{The Key Laboratory of Neuro-informatics \& Rehabilitation Engineering of Ministry of Civil Affairs, Xi'an, Shaanxi, 710049, P. R. China}

\author{Hai-Jun Zhou}
\affiliation{CAS Key Laboratory for Theoretical Physics, Institute of Theoretical Physics, Chinese Academy of Sciences, Beijing 100190, P.R. China}

\author{Jue Wang}
\email{juewang@xjtu.edu.cn}
\affiliation{The Key Laboratory of Biomedical Information Engineering of Ministry of Education, Institute of Health and Rehabilitation Science, School of Life Science and Technology, Xi'an Jiaotong University, Xi'an, Shaanxi, 710049, P. R. China}
\affiliation{National Engineering Research Center for Healthcare Devices. Guangzhou, Guangdong, 510500, P.R. China}
\affiliation{The Key Laboratory of Neuro-informatics \& Rehabilitation Engineering of Ministry of Civil Affairs, Xi'an, Shaanxi, 710049, P. R. China}

\author{Zi-Gang Huang}
\email{huangzg@xjtu.edu.cn}
\affiliation{The Key Laboratory of Biomedical Information Engineering of Ministry of Education, Institute of Health and Rehabilitation Science, School of Life Science and Technology, Xi'an Jiaotong University, Xi'an, Shaanxi, 710049, P. R. China}
\affiliation{National Engineering Research Center for Healthcare Devices. Guangzhou, Guangdong, 510500, P.R. China}
\affiliation{The Key Laboratory of Neuro-informatics \& Rehabilitation Engineering of Ministry of Civil Affairs, Xi'an, Shaanxi, 710049, P. R. China}

\author{Ying-Cheng Lai}
\affiliation{School of Electrical, Computer and Energy Engineering, Arizona State University, Tempe, AZ 85287, USA}
\affiliation{Department of Physics, Arizona State University, Tempe, Arizona 85287, USA}

\date{\today}

\begin{abstract}

The dynamical mechanism underlying the processes of anesthesia-induced loss of consciousness and recovery is key to gaining insights into the working of the nervous system. Previous experiments revealed an asymmetry between neural signals during the anesthesia and recovery processes. Here we obtain experimental evidence for the hysteresis loop and articulate the dynamical mechanism based on percolation on multilayer complex networks with self-similarity. Model analysis reveals that, during anesthesia, the network is able to maintain its neural pathways despite the loss of a substantial fraction of the edges. A predictive and potentially testable result is that, in the forward process of anesthesia, the average shortest path and the clustering coefficient of the neural network are markedly smaller than those associated with the recovery process. This suggests that the network strives to maintain certain neurological functions by adapting to a relatively more compact structure in response to anesthesia.

\end{abstract}

\maketitle

\section{Introduction} \label{sec:intro}

The origin of consciousness is an unsolved mystery in
nature~\cite{KMBT:2016,T:2008,M:2018,MWGP:2017,T:2017,EMWV:2016}. To
understand the neural physics of consciousness remains a challenging
problem, requiring interdisciplinary efforts among researchers from disciplines
such as neuroscience, physics, nonlinear dynamics and complex systems.
Anesthesia-induced loss of consciousness and the possible recovery open a door
to probing into the neural, physical, and dynamical mechanisms of
consciousness~\cite{H:2006,HM:2016,P:2013,AHT:2008}. Previous experimental
studies provided evidence for the existence of a hysteresis
phenomenon underlying the dynamics of loss and recovery of consciousness
induced by anesthesia~\cite{VVVCB:2006,FSMHMPJTESK:2010,VBCSSS:2012,KMML:2018}.
For example, signatures of a hysteresis were observed in behavioral
experiments on mice and drosophila subject to injection of
anesthetic and it was found that pharmacokinetics alone were not
sufficient to explain the emergence of the hysteresis~\cite{FSMHMPJTESK:2010}.
The concept of {\it neural inertia} was then proposed~\cite{FSMHMPJTESK:2010}.
Neural inertia and hysteresis in humans were investigated through behavior
indices~\cite{SCTRCCOC:2018,FCVFKMNA:2020} such as moments of loss and recovery
of responsiveness, as well as through EEG measurements~\cite{WSHJT:2017},
providing guidance to clinical anesthesia practices~\cite{STM:2019}.
In the work suggesting neural inertia and hysteresis in humans based on the
slow-wave activity in EEG through the dose-response relationship associated
with induction and emergence~\cite{WSHJT:2017}, saturation in the slow-wave
activity during anesthesia was found to depend on the age of the subject. This
indicated that neural inertia and hysteresis might have a neurobiological
basis in terms of the number of neurons and the synaptic density which
typically deteriorate with age.

In the anesthesiology literature, hysteresis is an established phenomenon
describing the changes in the patient EEG patterns as a function of drug
concentration, which can be explained by the pharmacokinetic-pharmacodynamic
(PKPD) model to some level. A source of debate is the extent to which the
hysteresis loop can (or cannot) be nulled out via appropriate adjustments to
the PKPD model used to compensate for delays between the measurement site
(either blood concentration for intravenous drugs such as propofol, or
end-tidal lung concentration for inhalational drugs such as isoflurane) and
the effect site, i.e., the brain. There has been extensive work on direct
observation of the hysteresis effect by
Kuizenga \emph{et al.}~\cite{KKH:1998,KWK:2001,KPWK:2001}, which
suggests that the PKPD model is likely an oversimplification of the complex
processes involved in generating the hysteresis effect in human anesthesia.
The hysteresis effect is thus likely to have additional causes.
In parallel, there were modeling efforts to understand the experimental
results. For example, a theory based on phase transition was proposed by 
Steyn-Ross \emph{et al.}~\cite{SSSL:1999,SSSW:2001,SSS:2004}, providing
an initial explanation of both biphasic excitation and hysteresis from the
perspective of neuronal populations.
Later, a percolation model based on reduced stochastic dynamics was proposed
to describe the mechanism of general anesthesia~\cite{ZMTX:2015}. Quite
recently, a multistate Markov chain model was introduced to interpret neural
inertia~\cite{PH:2018}.

The experimental evidence supporting the concept of neural inertia was
based on relatively indirect measurements, e.g., behavior indices or EEG,
of the neural system through noninvasive detection. From the perspective
of modeling, there is still a lack of understanding of neural inertia and the
corresponding hysteresis from a network perspective. The purpose of this paper
is to address the two issues. In particular, we have carried out animal
experiments and obtained evidence that a hysteresis loop arises unequivocally
in the processes of anesthetic administration and recovery after excluding
pharmacokinetic interference effects. Our result is based on the local field
potential (LFP) data that represent direct measurement of the neural system.
Our measurements have revealed that the depth of anesthesia is state dependent.
Motivated by the fact that state dependence is common in complex dynamical
systems~\cite{KP:2012,BIJPB:2000,HTD:2014} and by the existent theoretical
framework of modeling general anesthesia as a first-order phase transition
in the cortex~\cite{SSS:2004} in which the hysteresis is associated with state
dependence, we develop a complex-network based dynamical mechanism to
probe into the origin of the hysteresis phenomenon. The class of networks
we construct belongs to multilayer complex networks with self-similarity,
which involve the interactions of hierarchical units in the neural system
responsible for anesthesia and recovery. Computations reveal that, during
anesthesia, the network is able to maintain its neural pathways despite the
loss of a substantial fraction of the edges. A predictive and potentially
experimentally testable result from our network model is that, for a given
anesthesia level during the forward process, the corresponding characteristics
of the network, such as the clustering coefficient and average degree, can be
different from those in the recovery process. A biological implication is
that the network strives to maintain certain neurological functions by
adapting to a relatively more compact structure in response to dramatic
external disturbances such as anesthesia --- possibly a survival strategy
naturally gained during evolution of the nervous system. While the hysteresis
phenomenon arising in the anesthesia-recovery cycle can be explained at the
population level of neurons, our paper presents an alternative approach to
understanding the phenomenon from a network perspective.

\section{Experimental procedure and results} \label{sec:experiment}

Our experiments were conducted on two male six to eight week old C57BL/6 mice
at the time of surgery. Animals were housed in a standard environment on
a 12/12-h light/dark cycle with light on at 07:00, and they were allowed
\emph{ad libitum} access to water and food. The use and care of animals followed
the guidelines of the Xi'an Jiaotong University Animal Research Advisory
Committee, and all efforts were made to minimize animal suffering.

Mice were anesthetized using isoflurane (4$\%$ for induction, 2$\%$ for
maintenance in 100$\%$ O$_2$ during surgical procedures) and were placed in
a stereotaxic apparatus. A small animal heating pad was used for the
maintenance of body temperature at $37.0\pm 1.0^{\circ}\mbox{C}$.
Microelectrodes of impedance 0.5M$\Omega$ were implanted on the right medial
prefrontal cortex (1.7\mbox{mm} anterior to bregma, 0.4\mbox{mm} lateral to
the midline, and 1.5\mbox{mm} below the brain surface), which were used for
LFP recordings. For all the recordings, two screws implanted
in the occipital bone above the cerebellum were used as the ground and
reference. The whole implant was fixed with glass ionomer dental cement.

\begin{figure}[ht!]
\centering
\includegraphics[width=\linewidth]{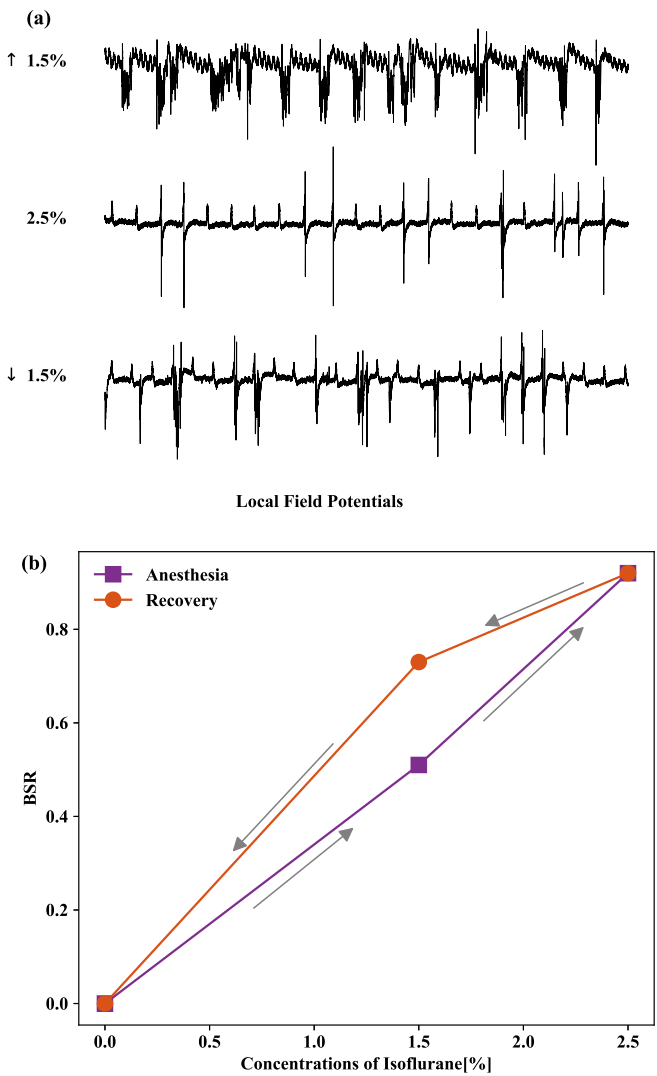}
\caption{ Representative experimental time series and direct evidence for the
emergence of a hysteresis loop. (a) Typical time series of mouse LFPs of 60s
during anesthesia and recovery. The data correspond to isoflurane concentration
rising to 1.5$\%$ (top), 2.5$\%$ (middle), and down to 1.5$\%$ (bottom),
respectively. (b) The corresponding hysteresis loop, where the ordinate is the
burst suppression ratio of mouse LFPs and the abscissa is the isoflurane
concentration. The arrows indicate the direction of the change in the
isoflurane concentration.}
\label{fig:Experiment}
\end{figure}

After surgery, all animals were allowed to recover for three to six days before
undergoing LFP recordings. LFPs were recorded extracellularly using a
128-channel data acquisition system (Cerebus, Blackrock Microsystems,
Salt Lake City, USA). Animals were placed in the induction chamber and
connected to the data recording cables, and oxygen was administered. LFP
recordings started 10 min before the induction of anesthesia. Isoflurane
concentrations were then stabilized at 1.5, 2.5, 1.5, and 0$\%$,
each for 20 min, to allow equilibration between inspired and end-tidal
concentration and thus eliminate the pharmacokinetic interference.
The time duration of 20 min
is sufficient to ensure the equilibration between the inspired and end-tidal
concentrations to eliminate the pharmacokinetics interference factors.
The burst suppression ratio (BSR) values shown in F\ref{fig:Experiment}(b) were obtained from a
5-min time interval after a 20-min transient period so as to avoid
the effect of unstable drug density during the transient phase of
pharmacokinetics. More specifically, the time series of LFPs used to calculate
the BSR value in F\ref{fig:Experiment}(b) were recorded when the effect
of isoflurane to the neural system already approaches equilibrium after
20 min, as verified through the statistical behaviors of the signals.
LFPs were collected at a sampling frequency of 1kHz, amplified
(300$\times$), and band-pass filtered (0.3-500Hz).

F\ref{fig:Experiment}(a) shows the measured LFPs from one of the mice
during anesthesia and recovery. In the rising and falling phases of
anesthetic concentration, the BSR~\cite{ICCZNZM:2009,VS:1998,HI:2007,KA:2007,CKBNMS:2014} of the LFPs is
different at the same level of concentration, signifying a hysteresis,
where BSR is defined as the fraction of the duration of the suppression
state in the total time interval (a suppression state is defined as the
LFP signal being between -0.1 and 0.1\mbox{mV} for a continuous time of at
least 100\mbox{ms}~\cite{VS:1998}). As shown in F\ref{fig:Experiment}(b),
the LFP during the falling phase has a larger burst suppression
ratio, suggesting that the mouse is still in deep anesthesia. The
hysteresis phenomenon shown in F\ref{fig:Experiment} cannot be fully
interpreted by pharmacokinetics, leading to the introduction of the concept
{\it neural inertia}~\cite{FSMHMPJTESK:2010}. Measurements of the second
mouse gave essentially the same result.

\begin{figure}[ht!]
\includegraphics[width=\linewidth]{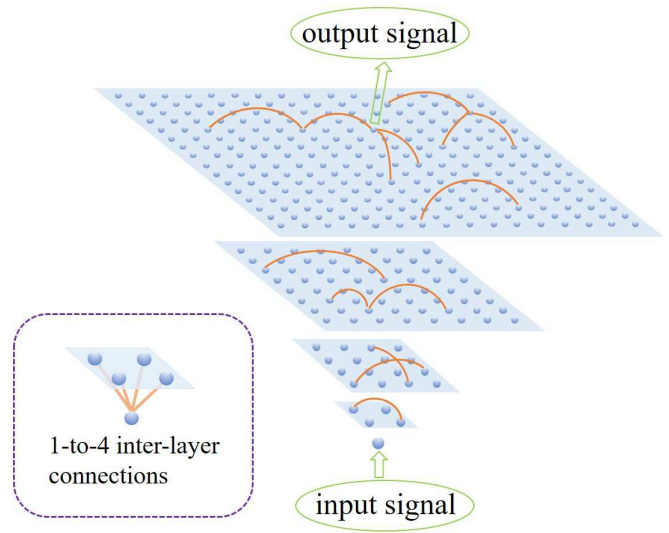}
\caption{Illustration of the proposed network model. The whole system has a
hierarchical multilayer structure. Each node in a lower layer is connected
 to four nodes in the adjacent upper layer (as shown in the dashed box).
Within each layer, the network can be scale free, small world, or random.
The input signal is applied to the root node and the output signal is taken
at a randomly selected node in the top layer.}
\label{fig:Network}
\end{figure}

\section{Model of the dynamical brain neural network for anesthesia and recovery}

\subsection{Model construction}

To interpret the experimental results, we develop a dynamical network model
based on physical considerations of the neural network structure and the
dynamics underlying the processes of anesthesia and recovery. For convenience,
we call the process of applying anesthetic ``forward'' and that of recovery
``backward.'' Structure-wise, in the two processes, many neural units are
involved. Regarding each neural unit as a node and the connection strength
between a pair of units as a weighted edge, we construct a class of multilayer
brain neural networks~\cite{ZMTX:2015}. An example is shown in
F\ref{fig:Network}, where the network consists of five layers with a
self-similar, fractal-like structure. The first (bottom) layer has only one
node, the root node, that serves as the modulator of the brain (thalamus) and
provides the driving force for the whole brain. For the other layers, in terms
of the possible structures of neuron connections, we consider general complex
network structures such as the scale-free~\cite{AB:2002},
small-world~\cite{WS:1998}, or random~\cite{ER:1959}
topology, to describe the nodal connections within each layer. The input
stimulation signal is applied to the root node and the corresponding output
signal is taken at a randomly selected node in the top layer.

We construct the network dynamics model based on the following intuitive
reasoning. In a typical setting, in the forward process, as the amount of
administrated anesthetic is increased, loss of consciousness occurs almost
instantaneously at a certain moment rather than gradually. Likewise, during
the backward (recovery) process, awakening tends to occur abruptly. The
empirical observation suggests that both the forward and backward processes
can be described by percolation dynamics on the underlying complex neural
network~\cite{ADS:2009,S:2015,ZMTX:2015}, corresponding to progressive edge
removal from and addition into the network, respectively. During the forward
process, failures of connections between the neural units occur as a result
of the action of anesthetic, depriving the units and those connected to them
of the ability to transmit and integrate information. Intuitively, the
connections among the more active nodes are more difficult to break than those
among less active nodes, which correspond to the larger or smaller degrees of nodes from the aspects of topology.
We thus assume that, at a certain anesthesia level,
the connection failure between a pair of nodes is random with a probability
determined by the degrees of both nodes. Effectively, there is mutual
maintenance among the hub units to sustain the vital neural connections in
the network. Neurological experiments indeed indicated that the normal
release of neurons is associated with maintenance and consolidation of the
synaptic connections and thus has a positive effect on the formation and
stable survival of the entire neural network~\cite{PGLL:2013}. More active
neurons with a higher firing rate typically have a large in-degree and thus
are more likely to form and maintain their connections with other
nodes~\cite{BP:1998}. Similarly, during the backward process, edges are
gradually restored in the network. For a pair of nodes, the probability of
recovery is determined by the current connectivity of the nodes. Consequently,
the connections between large-degree nodes in the presently ``broken'' network
are easier to recover but these nodes may not appear as the potential hub
nodes, especially in the early stage of the recovery process.

The physical reasoning supporting our construction of the dynamical neural
network suggests the following breaking and recovery probabilities associated
with an edge of end nodes $i$ and $j$:
$p_{ij}^b=1-(S_i+S_j)/(S_{\mathrm{max}}+S_{\mathrm{submax}})$ and
$p_{ij}^c=(S_i+S_j)/(S_{\mathrm{max}}+S_{\mathrm{submax}})$, respectively,
where $S_i=\sum_kw_{ik}$ characterizes the present strength of node $i$ and
$S_{\mathrm{max}}$ and $S_{\mathrm{submax}}$ are the maximum and the
second largest strengths at the present moment in the network, respectively.
The forward and reverse edge weights of the network are equal:
$w_{ij}=w_{ji}$. The level of anesthetic can be characterized by the failure
ratio $\rho$ --- the fraction of failed edges.

\begin{figure}[ht!]
\includegraphics[width=\linewidth]{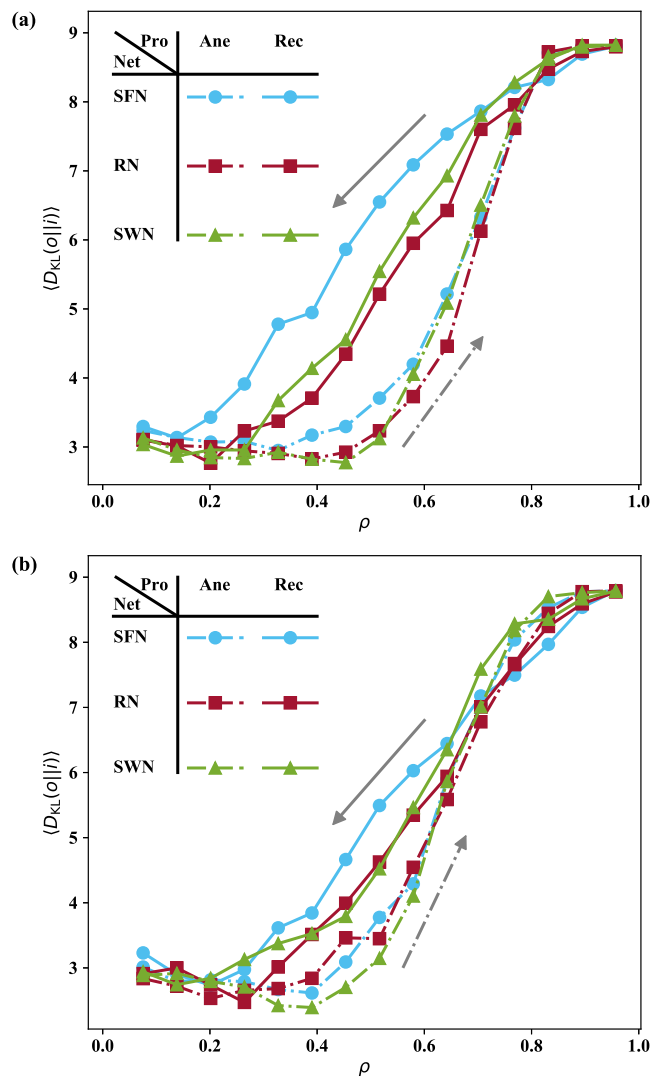}
\caption{Emergence of a hysteresis loop in the network structure associated
with the forward and backward processes. The hysteresis is revealed by the
Kullback-Leibler divergence defined as the ratio of the value distribution
of the output signal to that of the input signal vs the failure ratio
$\rho$ for (a) fixed edge weight $w=1$ and (b) uniformly distributed random
edge weight $w\sim U(0,2)$. Each data point is the result of averaging over
$100$ statistically independent realizations. For a single anesthesia-recovery
process, the transition from consciousness to unconsciousness is abrupt, and
vice versa. The input is a Gaussian signal
$\mathbb{N}(10 + \sin{(2\pi t/100)}, 1)$ with time-varying mean.
Results from three types of network models are included: scale-free
networks (SFNs), random networks (RNs), and small-world networks (SWNs).
The abbreviations ``Pro'', ``Net'', ``Ane'' and ``Rec'' stand for process,
network, anesthesia, and recovery, respectively. The same abbreviations are
used in subsequent figures.}	

\label{fig:Signal}
\end{figure}

\subsection{Emergence of hysteresis loops in various aspects of the
dynamical neural network}

Our main point is that the experimentally
observed hysteresis in F\ref{fig:Experiment} can be understood as the
result of simultaneous emergence of hysteresis loops in the characterizing
quantities of the dynamical network, which is established through a
systematic analysis of the dynamical network.

We first demonstrate the emergence of a hysteresis in the network structure.
We use the information transmission capability to measure the awareness
of the brain's nervous system, which can be characterized by the relative
entropy, also known as Kullback-Leibler divergence~\cite{KL:1951},
between the value distributions of the input and output signals:
$D_{\mathrm{KL}}(o||i) = \sum_{\alpha}
p_{\alpha}(o)\log{[p_{\alpha}(o)/p_{\alpha}(i)]}$,
where $p_{\alpha}(i)$ and $p_{\alpha}(o)$ are the probabilities for the
input and output signals to take on the value $\alpha$, respectively,
and $\alpha$ is calculated by the bin average. F\ref{fig:Signal} shows
the $D_{\mathrm{KL}}(o||i)$ curves for the forward
and backward processes. It can be seen that the two curves form a hysteresis
loop enclosing a finite area, indicating that the mechanism of mutual
maintenance between the network nodes can result in quite different network
structures for the forward and backward processes at the same anesthesia
level, especially at the intermediate levels. In general,
different network structures lead to different information transmission
capabilities and further to the experimentally observed hysteresis in the
burst suppression ratio of LFPs as demonstrated in F\ref{fig:Experiment}.
The hysteresis in the network structure is robust, as it arises regardless of
the topology of the network layers. Nonetheless, for randomly distributed
edge weights, the area of the hysteresis loop is smaller than that for fixed
weights.

\begin{figure}[ht!]
\centering
\includegraphics[width=\linewidth]{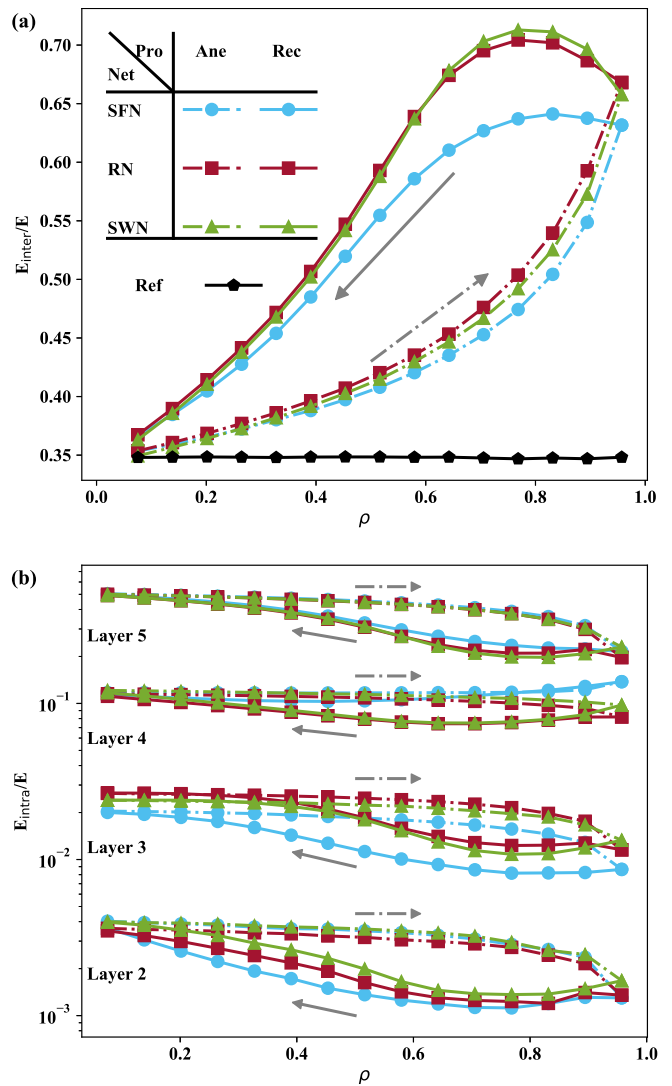}
\caption{Hysteresis in the network connectivity. (a) Interlayer edge ratio
vs the failure ratio $\rho$, where the black dotted line is the baseline
with random connection (or removal).
The abbreviation ``Ref'' stands for reference.
(b) Intralayer edge ratio vs $\rho$, where the dashed and solid curves
correspond to the forward and backward processes, respectively.}
\label{fig:Connectivity}
\end{figure}

We next demonstrate the emergence of another type of hysteresis, one manifested
as the difference in the network connectivity between the forward and backward
processes. F\ref{fig:Connectivity}(a) and F\ref{fig:Connectivity}(b)
show $E_{\mathrm{inter}}/E$ and $E_{\mathrm{intra}}/E$ for the forward and
backward processes versus the failure ratio $\rho$, where $E_{\mathrm{inter}}$,
$E_{\mathrm{intra}}$, and $E$ are the numbers of interlayer and intralayer
edges, and all edges in the network, respectively. In the forward process,
$E_{\mathrm{intra}}$ decreases rapidly with the deepening of anesthesia
since a large number of fragile edges with small degree ends are typically
intralayer edges, while $E_{\mathrm{inter}}$ decreases quite slowly. As a
result, the ratio $E_{\mathrm{inter}}/E$ increases monotonically with $\rho$
while $E_{\mathrm{intra}}/E$ decreases monotonically, as indicated
by the dashed curves in F\ref{fig:Connectivity}(a) and
F\ref{fig:Connectivity}(b), respectively. For the backward process where edges
are gradually restored in the network, a pair of nodes with large degree values
are more likely to be connected, and many such potential links are of the
interlayer type to connect the hub nodes across the layers. As the value of
$\rho$ gradually decreases due to continuous reduction in the anesthesia level,
$E_{\mathrm{inter}}$ increases rapidly but $E_{\mathrm{intra}}$ exhibits a
slow increase at first. However, the potential interlayer edges are not
many in comparison with the total number of possible edges in the network.
When most of the interlayer edges have been restored, the continued increase
in $E_{\mathrm{inter}}$ tends to slow down, leading to a rapid increase in
$E_{\mathrm{intra}}$. As a result, during the entire backward process,
$E_{\mathrm{inter}}/E$ first rises and then falls, as indicated by the solid
curves in F\ref{fig:Connectivity}(a) (from right to left), while
$E_{\mathrm{intra}}/E$ first falls and then rises, as revealed by the solid
curves in F\ref{fig:Connectivity}(b) (from right to left). Mutual
maintenance between nodes causes the forward and backward processes to have
different edge-connecting dynamics, leading to different network structures
for the two processes at the same anesthesia level (medium $\rho$). During
the entire backward process, the restoration of edges is comparatively
random, and the small cliques formed by these restored edges are not well
organized to support percolation between thalamus and cortex. The underlying
dynamics in the backward process are thus not compatible with the requirement
of consciousness recovering. The end result is a hysteresis that emerges in
the generation of network pathways between the forward and backward processes.

\begin{figure}[ht!]
\centering
\includegraphics[width=\linewidth]{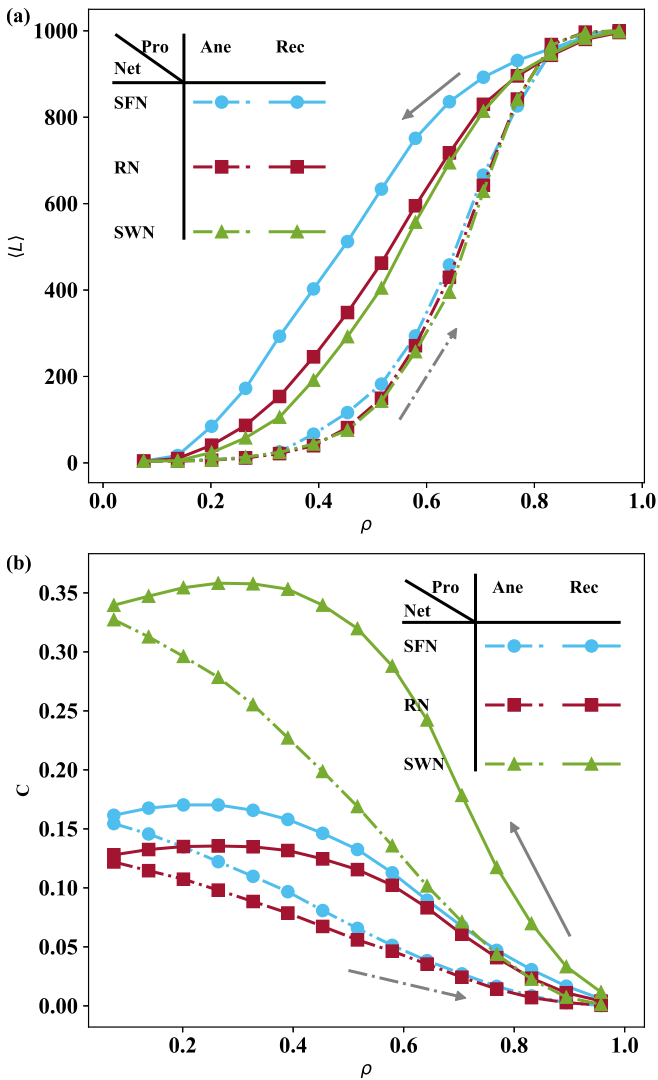}
\caption{ Hysteresis in the network characterizing quantities.
(a) Averaged shortest path $\langle L\rangle$ from the input node to output
node vs the failure ratio $\rho$. If there is no path between any pair
of nodes, we set $\langle L\rangle = 1000$, a value on the order of magnitude
of the total number of edges in the network. (b) Clustering coefficient $C$
vs $\rho$. Both $\langle L\rangle$ and $C$ exhibit a hysteresis.}
\label{fig:Characterization}
\end{figure}

The hysteresis in the network structure manifests itself as hystereses in the
characterizing quantities of the network, such as the average shortest path
$\langle L\rangle$ from the input to the output node and the clustering
coefficient $C$, as shown in F\ref{fig:Characterization}(a) and
F\ref{fig:Characterization}(b), respectively. For intermediate values of the
$\rho$, the values of $\langle L\rangle$ and $C$ associated with
the forward process are significantly lower than those with the backward
process, indicating that, under the same anesthesia level, the underlying
neural network of the forward process transmits information more rapidly
with more straightforward and less redundant network connectivity than that
with the backward process. Biologically, this may be interpreted as a kind
of functional maintenance that the brain nervous system struggles to achieve
in response to threats.

\section{Discussion} \label{sec:discussion}

To summarize, we have obtained experimental evidence of the emergence of a
hysteresis loop underlying the anesthesia and recovery processes. To preclude
possible interference from pharmacokinetics, we have carried out direct
measurement of the local field potentials of the neural system through
invasive detection. We have developed a physical understanding based on
multilayer, self-similar networks with a complex topology. The specific
neurophysiological mechanism of the hysteresis is unknown at the present, but
our model suggests that the mutual maintenance between certain units of the
nervous system may play an important role in generating the hysteresis. The
mutual interactions, due to the heterogeneous connectivity of the nervous
system, can result in a significant difference in the dynamical evolution of
neural connections between the anesthesia and recovery processes. The complex
brain neural network is the result of long term evolution and has embodied
some kind of efficiency and adaptability with functional robustness in response
to external disturbances or threats~\cite{VSKP:2009,TWV:2013,HG:2017,ZY:2018}.
One of our findings is that, in the forward process of anesthesia, the average
shortest path and the clustering coefficient of the neural network are
markedly smaller than those associated with the recovery process
(F\ref{fig:Characterization}).
This suggests that the network strives to
maintain certain neurological functions by adapting to a relatively more
compact structure in response to anesthesia --- a kind of dramatic external
disturbance. This represents a survival strategy naturally gained as the
result of evolution of the nervous system.
Additionally, our network model predicts that the structural characteristics
of the underlying network, such as the clustering coefficient and average
degree, can be quite different during the anesthesia and recovery processes.
An experimental test of this prediction is called for.

There has been experimental work indicating that noise tends to attenuate the
hysteresis associated with the anesthesia-recovery process~~\cite{PH:2018},
which can actually be explained with our network approach. In particular,
hysteresis is a phenomenon in multistable dynamical systems going
through state transitions. Noise can enhance the probability for a state
transition to occur, thereby weakening the hysteresis effect. From the point
of view of either neuronal populations or network, the anesthesia-recovery
process can be viewed as the transitions of the nervous system between two
stable states, so the underlying system is effectively bistable. In the
forward process, the robust connections among the large degree nodes mean
that the main pathway between the input and output nodes can always be
maintained, regardless of perturbations. In this sense, noise has little
effect on the forward process. However, noise can have a significant effect
on the backward process, which can be seen by noting that, in the nervous
system, noise typically manifests itself as some kind of remote excitatory
stimulus that specifically enhances the excitability of the neurons and
accordingly synaptic connections. In our network model, conceptually the
role of noise in enhancing neuronal connections is equivalent to adding extra
edges into the network. The extra edges significantly increase the probability
of the emergence of pathways between the input and output nodes, advancing the
occurrence of a fully connected network state. The global effect of noise is
thus to weaken the hysteresis.

\begin{acknowledgments}

We thank Prof. Celso Grebogi and ChangSong Zhou for helpful discussions. This work was supported by the National Natural Science Foundation of China 
(No. 11975178, No. 11647052, and No. 6143101), the Open Project of State Key Laboratory of Cognitive Neuroscience and Learning (No. CNLYB1802), and the Project Supported by Natural Science Basic Research Plan in Shaanxi Province of China (No. 2020JM-058). ZGH acknowledges the support of the K. C. Wong Education Foundation and the Fundamental Research Funds for the Central Universities (xjtu2019). YCL is supported by the Office of Naval Research through Grant No. N00014-16-1-2828.

\emph{Attributions.} C.-W.S. and L.Z. contributed equally to this work.

\end{acknowledgments}


%
\end{document}